\RequirePackage{fix-cm}

\documentclass[smallextended, authoryear]{svjour3}      
\smartqed  

\usepackage{graphicx}
\usepackage{natbib}

\usepackage{algorithmic}
\usepackage{graphicx}
\usepackage{textcomp}
\usepackage{xcolor}
\usepackage{amsmath}
\usepackage{makecell}
\usepackage{subcaption}
\usepackage{listings}
\usepackage{fancyhdr}
\usepackage[dvipsnames]{xcolor}
\usepackage{multirow}
\usepackage{soul}
\definecolor{mygreen}{HTML}{2DA44E}
\usepackage{balance}
\usepackage{placeins}
\usepackage{enumitem}

\usepackage{tcolorbox} 

\newcommand{\rqboxc}[1]{\begin{tcolorbox}[left=3pt,right=3pt,top=3pt,bottom=3pt,colback=gray!5,colframe=gray!40!black,before skip=10pt,after skip=10pt]#1\end{tcolorbox}}

\newcommand{\phead}[1]{\vspace{1mm} \noindent {\bf #1}}

\newcommand{\iuhead}[1]{\vspace{1mm} \noindent {\ul {\textit {#1}}}}

\usepackage{booktabs}

\definecolor{dkgreen}{rgb}{0,0.6,0}
\definecolor{gray}{rgb}{0.5,0.5,0.5}
\definecolor{mauve}{rgb}{0.58,0,0.82}
\definecolor{darkgreen}{rgb}{0.01, 0.75, 0.24}
\definecolor{red}{RGB}{200,0,0}
\definecolor{blue}{RGB}{0,70,200}

\usepackage{fvextra}
\fvset{
fontsize=\scriptsize,
breaklines=true,
breakanywhere=true
}
\usepackage{tabularx}

\usepackage{url}

\definecolor{javared}{rgb}{0.6,0,0} 
\definecolor{javagreen}{rgb}{0.25,0.5,0.35} 
\definecolor{javapurple}{rgb}{0.5,0,0.35} 
\definecolor{javadocblue}{rgb}{0.25,0.35,0.75} 
\definecolor{lightred}{HTML}{FFEBEE}

\lstdefinestyle{mystyle}{
  frame=single,
  xleftmargin=4pt,
  xrightmargin=4pt,
  abovecaptionskip=2pt,
  belowcaptionskip=0pt,
  captionpos=b,
  escapeinside={*‘}{’*},
  tabsize=4,
  emphstyle={\bf},
  basicstyle=\footnotesize\ttfamily,
  keywordstyle=\color{javapurple}\bfseries,
  stringstyle=\color{javared},
  commentstyle=\color{javagreen},
  morecomment=[s][\color{javadocblue}]{/**}{*/},
  showspaces=false,
  columns=flexible,
  showstringspaces=false,
  morecomment=[l]{//},
  breaklines=true
}

\usepackage{colortbl}

\definecolor{topone}{HTML}{C4483F}
\definecolor{toptwo}{HTML}{EFA9AD}
\definecolor{topthree}{HTML}{F8CACE}
\definecolor{grouphead}{HTML}{F2F4F8}
\definecolor{lightgray}{gray}{0.95}

\newcommand{\codeoperation}[1]{\textcolor{javapurple}{\bfseries #1}}

\usepackage{hyperref}
\begin{document}

\title{Rethinking Code Performance Benchmarks for LLMs}

\author{
Nhat Minh Le \and
Yisen Xu \and
Zhijie Wang \and
Tse-Hsun (Peter) Chen
}

\institute{
Nhat Minh Le \at
Software Performance, Analysis, and Reliability (SPEAR) Lab, Concordia University, Montreal, Quebec, Canada \\
\email{nhatminh.le@mail.concordia.ca}
\and
Yisen Xu \at
Software Performance, Analysis, and Reliability (SPEAR) Lab, Concordia University, Montreal, Quebec, Canada \\
\email{yisen.xu@mail.concordia.ca}
\and
Zhijie Wang \at
Concordia University, Montreal, Quebec, Canada \\
\email{zhijie.wang@concordia.ca}
\and
Tse-Hsun (Peter) Chen \at
Software Performance, Analysis, and Reliability (SPEAR) Lab, Concordia University, Montreal, Quebec, Canada \\
\email{peterc@encs.concordia.ca}
}

\date{Received: date / Accepted: date}

\maketitle

\begin{abstract}

Many function-level performance benchmarks have been proposed to evaluate whether large language models (LLMs) can generate efficient programs. However, results on these benchmarks often show that LLM-generated implementations have little or no execution-time difference from canonical solutions. This observation motivates us to revisit these benchmarks and examine whether they are suitable for performance evaluation. In this paper, we revisit four popular benchmarks: EffiBench, Enamel, EvalPerf, and Mercury. We evaluate 1,538 tasks under more rigorous setting by running each task 30 times and assessing the runtime differences between the canonical solutions and benchmark-provided performant implementations with statistical testing. With the benchmark-provided test suites, only 6.11\% of the performant implementations are significantly faster than the canonical solutions. In a manual analysis of 308 non-significant tasks, 99 performant implementations contain no meaningful performance change, while 209 contain potential performance improvements that are not exposed by the original tests.

These results suggest that the main limitation is not only the evaluation method, but also the limited sufficiency of the benchmark-provided performance tests. To address this limitation, we propose an LLM-based multi-agent framework to generate performance-oriented tests that expose runtime differences more effectively than the original tests. The framework uses three separate agents to generate, diagnose, and repair deterministic tests that preserve functional correctness while better exposing performance differences. Across 1,345 benchmark tasks for which the original tests found no significant performance difference, tests generated by our framework with DeepSeek-v3.1 and GPT-4o reveal statistically significant improvements in 24.01\% and 25.43\% of the tasks, respectively, outperforming the SOTA LLM-based performance test generation method. 
Finally, we discuss the implications for future performance benchmark for LLM-generated code. In addition to using repeated execution and statistical testing to improve rigor, future work should consider selecting problems with meaningful opportunities for performance optimization rather than relying on overly simple tasks, constructing sufficiently challenging test cases that make runtime differences between implementations observable, and extending this line of evaluation from isolated function-level tasks to class-level or repository-level settings where performance bottlenecks may arise from interactions among multiple components. 
\end{abstract}


\section{Introduction}\label{sec:introduction}

Software performance has a long research history in the software engineering community~\citep{woodside2007future}. Decades of work have studied how to measure, diagnose~\citep{jin2012understanding,baltes2015navigate}, and improve runtime behavior through performance testing~\citep{vokolos1998performance, weyuker2000experience}, workload characterization~\citep{avritzer2002software}, performance debugging, benchmarking~\citep{mytkowicz2009producing, kalibera2013rigorous}, and statistically rigorous performance~\citep{georges2007statistically}. As software systems increasingly rely on automated code generation, this long-standing problem has re-emerged in a new setting: whether code produced by large language models (LLMs) is not only functionally correct but also computationally efficient.

A series of benchmarks for evaluating the correctness of LLM-generated code have been proposed at the function level~\citep{humaneval, mbpp, apps2021, ds10002023, li2022competition, jain2025livecodebench, zhuo2025bigcodebench}, class level~\citep{du2023classeval, cao2024javabench, rahman2025beyond, chen2026classeval}, or repository level~\citep{swebench2024, repobench2024, li2024deveval, le2025impacts, li2025fea}. 
Notably, function-level Python code generation benchmarks, such as HumanEval~\citep{humaneval} and MBPP~\citep{mbpp}, have become the dominant evaluation setting because individual functions naturally serve as self-contained, executable units with well-defined input-output specifications. Compared with class or repository-level generation, function-level tasks largely avoid confounding factors such as project organization, dependency management, or build configuration, allowing model performance to be attributed more directly to code synthesis ability. Furthermore, the generated functions can be reliably evaluated using unit tests, enabling scalable and reproducible comparison across models.

Nevertheless, correctness alone does not tell us whether generated code is computationally efficient. To address this gap, recent work has developed function-level Python program efficiency benchmarks, such as EffiBench~\citep{effibench}, Enamel~\citep{enamel}, EvalPerf~\citep{evalperf}, and Mercury~\citep{mercury}. Each task includes a canonical solution and a benchmark-provided performant implementation. The canonical solution serves as the baseline for evaluating LLM-generated implementations, whereas the performant implementation demonstrates that the task admits a more efficient solution. During evaluation, an LLM-generated implementation is first validated for correctness and, if correct, its execution time is compared against the canonical solution using the benchmark-provided test cases.
Yet, evaluations using these benchmarks often find that LLM-generated implementations show little runtime difference from, or remain slower than, the corresponding canonical implementations. 
For example, a recent study~\citep{islam2025evaluating} shows that even strong LLMs remain slower than human-written canonical reference solutions: on 298 commonly solved LeetCode tasks solved by all 20 models, the canonical reference code runs in 74.16 ms on average, while LLM-generated solutions range from 75.64 ms for DeepSeek-V3~\citep{deepseek_model} to 147.95 ms for GPT-4 Turbo~\citep{gpt4}. Before attributing such results solely to model limitations, we first ask whether the benchmarks can expose intended optimizations at all. Surprisingly, even benchmark-provided performant implementations often show little or no runtime improvement over the corresponding canonical solutions under the current evaluation settings. This suggests that the evaluation settings of these benchmarks may not fully capture meaningful execution-time differences.

Two factors can contribute to this problem. First, many benchmarks do not adopt statistically rigorous measurement procedures. Execution times are often collected from only a single run (e.g., Effibench~\citep{effibench} and Mercury~\citep{mercury}) or a small number of runs (e.g., 6 times in Enamel~\citep{enamel}), without repeated measurements and statistical significance testing to determine whether observed differences are reliable rather than the result of measurement noise. It remains unclear whether the reported performance differences between LLM-generated implementations and reference implementations reflect genuine efficiency improvements.
Second, many performance benchmarks are derived from existing code-generation benchmarks such as HumanEval~\citep{humaneval} and MBPP~\citep{mbpp}, where test cases are usually designed to assess functional correctness rather than execution efficiency. As a result, the benchmark-provided test inputs may not be sufficiently challenging to reveal meaningful execution-time differences between implementations. Implementations with substantially different algorithmic or optimization characteristics may appear to perform similarly under the existing benchmark’s evaluation setting.

These observations raise a fundamental question: \textit{do current benchmarks reliably distinguish genuinely efficient implementations from merely functionally correct ones?} In this paper, we revisit four function-level Python
performance benchmarks using a more statistically rigorous performance evaluation. We study 1,538 tasks from EffiBench~\citep{effibench}, Enamel~\citep{enamel}, EvalPerf~\citep{evalperf}, and Mercury~\citep{mercury}. For each task, we execute both the canonical solution and the performant implementation 30 times on the benchmark-provided
tests, providing sufficient observations for stable statistical analysis while maintaining a practical computational cost~\citep{ArcuriB11,zeng2019studying, su2019pinpointing, liao2025early, JangaliTALYS23}. 

Our re-evaluation shows that most benchmark-provided performant
implementations do not exhibit statistically significant ($p \geq 0.05$) execution-time improvements under the original test suites. Across 1,538 tasks, only 94 performant implementations (6.11\%) are significantly faster than their corresponding canonical solutions, while 1,444 implementations (93.89\%) are not statistically distinguishable from the canonical solutions.

To understand why these implementations remain statistically indistinguishable, we analyze the corresponding benchmark tasks. We sampled 308 tasks from the 1,444 non-significant tasks across the four benchmarks for manual analysis. The sample provides a 95\% confidence level with a 5\% margin of error. Two of the authors performed open coding (final Cohen’s Kappa: 0.93, indicating almost perfect agreement) and found two main reasons. First, 99 out of 308 performant implementations contain no meaningful performance change, such as refactoring or no code changes. The remaining 209 tasks contain algorithmic or data-structure changes that should plausibly improve runtime performance. However, the original test inputs do not stress the code paths where these changes matter. 
We further extended our analysis with LLM-as-a-Judge~\citep{ahmedDTP25,lillmjudge2024,zhangLWSMS024} over the remaining 1,136 non-significant tasks. We first confirm substantial agreement between human labelers and two LLMs (DeepSeek-V3.1 and GPT-40-2024-08-06) on the 308 manually labeled tasks, with Cohen’s Kappa scores of 0.72 and 0.75, respectively. We observe the same pattern at larger scale: DeepSeek-V3.1 and GPT-4o-2024-08-06 label 76.06\% and 73.94\% of these tasks, respectively, as changes with potential performance impact.

To generate test suites that can effectively reveal the performance differences, we develop a multi-agent test-generation framework that produces tests intended to preserve correctness while increasing the computational load on performance-critical paths. 
Our framework adopts a batch-based multi-agent workflow that iteratively generates, executes, diagnoses, and repairs performance-oriented tests before using the validated suite for performance evaluation. With the tests generated by our framework, more performant implementations become statistically distinguishable from their canonical solutions: among 1,345 previously non-significant tasks, DeepSeek-v3.1 and GPT-4o generated tests reveal statistically significant improvements in 323 (24.01\%) and 342 (25.43\%) tasks, respectively. When we re-evaluate implementations generated
by GPT-4o-mini~\citep{gpt_4o_mini}, Claude-Sonnet-4.5~\citep{claude_sonnet_4_5}, and Gemini-2.5-Flash~\citep{gemini_2_5_flash}, stronger tests reveal statistically significant performance improvements in 22.19\% of the evaluated cases. Overall, our findings suggest that conclusions drawn from existing function-level Python performance benchmarks should be interpreted with caution. Reliable evaluation of LLM-generated code efficiency requires both statistically rigorous execution measurement and test suites that sufficiently stress performance-critical behavior.

Our contributions are summarized as follows:

\begin{itemize}

\item We revisit four function-level Python performance benchmarks using repeated execution, statistical testing, and effect-size analysis.

\item We investigate why benchmark-provided performant implementations are often not distinguishable from canonical solutions, combining manual analysis with a validated LLM-as-a-Judge procedure.

\item We develop a multi-agent framework that generates, diagnoses, and repairs performance-oriented tests to better stress performance-critical behavior.

\item We demonstrate that stronger performance-oriented test suites can substantially alter conclusions about the efficiency of LLM-generated code.

\end{itemize}

\phead{Paper Organization.} Section~\ref{sec:background} introduces the background and motivation. Section~\ref{sec:studydesign} describes the study design and evaluation metrics. Section~\ref{sec:results} presents the results for each research question. Section~\ref{sec:discussion} discusses the findings and potential future research directions. Section~\ref{sec:related} reviews related work. Section~\ref{sec:threats} examines threats to validity. Finally, Section~\ref{sec:conclusion} concludes the paper.
\section{Background and Motivation}\label{sec:background}

In this section, we first introduce performance benchmark for LLM's code generation. Then we discuss the challenges that motivate our study. 

\subsection{Performance Benchmark for LLM's Code Generation}

A function-level performance benchmark evaluates whether an LLM can generate code that is both functionally correct and computationally efficient. Let a benchmark be a collection of tasks
$\mathcal{B}=\{\tau_1,\tau_2,\ldots,\tau_n\}$. Each task $\tau$ can be represented as
\[
\tau=(d_\tau, \mathcal{T}_\tau, c_\tau, p_\tau),
\]
where $d_\tau$ is the natural-language problem description, $\mathcal{T}_\tau$ is the benchmark-provided set of executable test cases, $c_\tau$ is the canonical solution, and $p_\tau$ is the benchmark-provided performant implementation. Each test case in $\mathcal{T}_\tau$ specifies the inputs and expected outputs used to verify functional correctness. The canonical solution serves as the baseline implementation for evaluating the runtime efficiency of LLM-generated code. The benchmark-provided performant implementation plays a different role: it is a reference artifact indicating that the task can be optimized and showing one possible optimized implementation.

Let $M$ denote the LLM-based code generator being evaluated. For each task, the benchmark asks $M$ to synthesize an implementation $g_\tau=M(d_\tau)$. 
Only implementations that pass all the test cases in $\mathcal{T}_\tau$ are considered for performance evaluation.

Let $\widehat{T}(f,\tau)$ denote the runtime reported by the benchmark after executing implementation $f$ on $\mathcal{T}_\tau$. 
For LLM-generated code, the runtime comparison is made against the canonical solution. After a generated implementation $g_\tau$ passes the functional tests, its runtime $\widehat{T}(g_\tau,\tau)$ is compared with the canonical runtime $\widehat{T}(c_\tau,\tau)$. The generated implementation is considered faster when its measured runtime is lower than the canonical runtime. The benchmark-provided performant implementation $p_\tau$ is not the target that generated code must strictly outperform.

Instead, $p_\tau$ is an optimized reference solution. Its role is to show that task $\tau$ has an intended optimization opportunity and to illustrate how the canonical solution can be improved. Therefore, the measured runtime difference between $p_\tau$ and $c_\tau$ should reflect a meaningful efficiency difference, with $p_\tau$ running faster than $c_\tau$. If even this comparison does not show a reliable difference, then the benchmark may not be able to reveal performance improvements in LLM-generated code either.

\subsection{Challenges in Building Performance Benchmarks for LLM-Generated Code}

We argue two major challenges exist when building reliable performance benchmarks for LLM-generated code.

\phead{Challenge 1: Runtime Measurement Noise.} Runtime measurements are inherently noisy. Even identical code executed on the same hardware can produce different runtimes across runs due to operating system scheduling, CPU cache state, and background process interference~\citep{chenR16noise,mogulB91contextswitch,tsafrirEFK05systemnoise}. The performance engineering community has long established that single-run comparisons are unreliable and commonly recommends at least 30 executions when comparing runtime performance~\citep{ArcuriB11}. Nevertheless, benchmarks such as EffiBench and Mercury execute each implementation only once per task, making their evaluations vulnerable to measurement noise. Enamel and EvalPerf repeat executions 6 and 5 times, respectively, which reduces measurement variability but still provides very limited support for rigorous statistical inference.

{
\setlength\fboxsep{0pt} 

\begin{samepage}
\begin{center}
\begin{minipage}[t]{0.48\linewidth}
\textbf{Canonical solution}
\begin{lstlisting}[language=Python, escapechar=!, numberstyle=\tiny\color{lightgray}]
def fib(self, n: int) -> int:
!{\colorbox{yellow!30}{    \codeoperation{if} n == 1: \codeoperation{return} 1}}!
!{\colorbox{yellow!30}{    \codeoperation{if} n <= 0: \codeoperation{return} 0}}!
    return self.fib(n-1) + self.fib(n-2)
\end{lstlisting}
\end{minipage}
\hfill
\begin{minipage}[t]{0.48\linewidth}
\textbf{Performant implementation}
\begin{lstlisting}[language=Python, escapechar=!, numberstyle=\tiny\color{lightgray}]
def fib(self, n: int) -> int:
!{\colorbox{yellow!30}{    \codeoperation{if} n == 0: \codeoperation{return} 0}}!
!{\colorbox{yellow!30}{    \codeoperation{elif} n == 1: \codeoperation{return} 1}}!
    return self.fib(n-1) + self.fib(n-2)
\end{lstlisting}
\end{minipage}
\captionsetup{type=lstlisting,hypcap=false}
\caption{Changes that do not have performance impact.}
\label{example_code}
\end{center}
\end{samepage}
}

On the other hand, benchmark conclusions can be misleading in the presence of measurement noise. Example~\ref{example_code} shows a task (Task ID: 509) from Mercury. The code snippet at the right is labeled by the benchmark as the more efficient implementation. However, it only reorders conditional statements and uses {\sf elif}, without changing the algorithm or its asymptotic runtime complexity. Consequently, neither implementation should be expected to exhibit a meaningful runtime advantage. The reported performance difference is therefore likely attributable to measurement noise rather than genuine efficiency gains. Indeed, when we executed both the canonical solution and the performant implementation 30 times, we observed no statistically significant runtime difference (Mann-Whitney U test p-value: 0.18, effect size: 0.1).

\phead{Challenge 2: Insufficient Test Inputs.} Even with nearly noise-free measurements, runtime comparisons are meaningful only when test inputs impose sufficient computational load to expose algorithmic differences. This is a fundamental challenge for current performance benchmarks, because many are derived from correctness-oriented code generation benchmarks such as HumanEval~\citep{humaneval} and MBPP~\citep{mbpp}. As a result, their test suites are designed to verify functional correctness rather than evaluate runtime efficiency. For example, suppose one implementation replaces an $O(n^3)$ algorithm with an $O(n^2)$ algorithm. When evaluated on a very small input (e.g., $n=4$), both implementations complete within microseconds, making their runtime difference negligible despite the asymptotic improvement. In such cases, the benchmark cannot expose the optimization because the test inputs never exercise the performance-critical behavior. As a result, genuinely faster implementations may appear no different from their baseline counterparts. This is not because the optimizations are ineffective, but because the benchmark test inputs are insufficiently demanding.

These challenges motivate us to first examine the reliability of existing function-level Python performance benchmarks. Before using these benchmarks to evaluate LLM-generated code, we ask whether the benchmark-provided performant implementations are distinguishable from the corresponding canonical solutions under the benchmark's own test cases and a more rigorous measurement procedure. If they are not, the benchmark cannot meaningfully assess whether generated implementations are truly more efficient. Instead, the observed results may be driven by measurement noise or insufficiently demanding test inputs rather than genuine performance differences, undermining conclusions about both benchmark quality and LLMs’ ability to generate efficient code.

\section{Study Design}\label{sec:studydesign}

\subsection{Benchmarks and Data}

We collect tasks from four performance-oriented function-level code generation benchmarks: EffiBench~\citep{effibench}, Enamel~\citep{enamel}, EvalPerf~\citep{evalperf}, and Mercury~\citep{mercury}. We collected a total of 1,538 tasks. Table~\ref{tab:overview} shows the distribution. For each task, we obtain two implementations: the \textit{canonical solution} and the \textit{benchmark-provided performant implementation}.

\begin{table}[htbp]
\centering
\caption{An overview of the benchmarks. \textit{Reported Runs} shows the number of repeated executions per task reported by the corresponding paper. }
\label{tab:overview}
\small
\begin{tabular}{l l r r}
\toprule
\textbf{Benchmark} &
\textbf{Task Source} &
\textbf{\#Tasks} &
\textbf{Reported Runs} \\
\midrule
EffiBench & LeetCode & 1,000 & 1 \\
Enamel    & HumanEval / HumanEval+ & 164 & 6 \\
EvalPerf  & HumanEval / MBPP & 118 & 5 \\
Mercury   & LeetCode & 256 & 1 \\
\midrule
\textbf{Total} & -- & 1,538 & -- \\
\bottomrule
\end{tabular}
\end{table}

The canonical solutions are provided by each benchmark's task source. For example, the canonical solutions in Enamel~\citep{enamel} and EvalPerf~\citep{evalperf} come from existing correctness-oriented datasets such as HumanEval~\citep{humaneval} and MBPP~\citep{mbpp}.
EffiBench derives canonical solutions from the corresponding LeetCode discussion forum, consisting of user-submitted solutions that have been accepted by the platform. In contrast, Mercury constructs canonical solutions from historical accepted submissions for each task. 

In terms of collecting the performant implementations, for EffiBench, the performant implementations are generated by GPT 3.5 Turbo in their paper. 
For Mercury~\citep{mercury}, performant implementations are derived by aggregating historical LeetCode submissions to construct a runtime distribution and identify faster solutions. 
In Enamel~\citep{enamel} and EvalPerf~\citep{evalperf}, the performant implementations are human-optimized versions of the corresponding canonical solutions.

\subsection{Execution Procedure and Statistical Analysis}

For each task, we execute both the canonical solution and the benchmark-provided performant implementation using the test inputs supplied by the benchmark. During each run, we record the execution time of the program.

To reduce the impact of transient system noise, such as operating system scheduling, CPU cache effects, and background processes~\citep{chenR16noise,mogulB91contextswitch,tsafrirEFK05systemnoise}, each implementation is executed 30 times while keeping inputs and environmental conditions constant. This follows prior performance measurement research that relies on repeated executions to reduce unreliable single-run comparisons~\citep{ArcuriB11,zeng2019studying, su2019pinpointing, liao2025early, JangaliTALYS23}. These repeated runs produce independent runtime samples for the canonical and performant implementations, allowing statistical analysis to assess whether observed runtime differences are statistically significant and not attributable to measurement noise.

\phead{Statistical Hypothesis Test.}
Following the stability-oriented practice used in prior software performance research~\citep{ArcuriB11}, we do not rely on aggregate execution-time summaries alone to decide whether an implementation is faster. Instead, for each task, we compare the full runtime samples collected from repeated executions of the canonical solution and the benchmark-provided performant implementation. Before applying any statistical test, we first check functional correctness and execution validity. If the benchmark-provided performant implementation fails to pass at least the same test cases as the canonical solution, or if any execution times out, we treat the task as not statistically significant for performance improvement. In other words, such tasks are assigned to the non-significant category, equivalent to failing the $p<0.05$ decision criterion. 

For tasks with valid executions, We use the Mann-Whitney U test~\citep{mann1947test}, a non-parametric test that does not assume normally distributed runtimes and is suitable for comparing two independent runtime samples. The null hypothesis is that the performant implementation is not faster than the canonical solution, while the alternative hypothesis is that its runtime distribution is shifted toward lower execution times. We consider a task statistically significant only when the test yields $p<0.05$. To analyze performance differences across tasks at the benchmark level, we compute one average runtime for each implementation on each task and apply a one-sided Wilcoxon signed-rank test~\citep{wilcoxon1945individual,sidney1957nonparametric} to these paired per-task averages. This benchmark-level test is non-parametric and is well suited for paired measurements across tasks~\citep{demvsar2006statistical}, allowing us to examine whether performant implementations are consistently faster than canonical solutions within each benchmark.

\phead{Effect Size.}
To avoid overinterpreting statistically significant but practically negligible differences~\citep{kampenes2007systematic}, we measure effect size using Cliff's delta~\citep{cliff1993dominance}. We interpret the magnitude of $\delta$ using standard thresholds: negligible ($\delta < 0.147$), small ($0.147 \le \delta < 0.33$), medium ($0.33 \le \delta < 0.474$), and large ($\delta \ge 0.474$).

\subsection{Experimental Environment}

All experiments were conducted on a desktop machine equipped with an Intel® Core™ i5-8500 CPU @ 3.00 GHz (6 physical cores, 6 logical threads, 9 MB L3 cache). To reduce runtime measurement noise, we ran only one benchmark task at a time on the machine and ensured that no other programs or processes were running during test execution. We used the same execution timeout for all evaluated implementations: each test run was terminated if it exceeded 300 seconds.
For LLM-as-a-Judge in RQ2 and test generation in RQ3, we used DeepSeek-V3.1~\citep{deepseek_chat_2025} and GPT-4o (gpt-4o-2024-08-06)~\citep{openai_chatgpt_4o_2024_08_06}. These models are selected for their strong reasoning capabilities and accessible APIs. For all LLM calls, we set the temperature to 0 to reduce the randomness and improve replicability.
\section{Results}\label{sec:results}

\subsection{RQ1: Can Existing Benchmarks Reliably Expose Performance Differences?}

\phead{Motivation.}
The reliability of existing benchmarks remains unclear. This research question aims to establish a fair and robust runtime performance measurement framework and rigorously evaluate whether the benchmark-provided performant implementations are significantly faster than the corresponding canonical solutions.

\phead{Approach.} 
For each of the 1,538 tasks, we execute both the canonical solution and the performant implementation 30 times using the benchmark-provided test suites. We record the runtime of each run and compute the mean for both the canonical solution and the performant implementation. For each task, we further apply Mann-Whitney U test and compute Cliff's delta on the runtime measurements.

\begin{table}[htbp]
\centering
\caption{Mann-Whitney U test results across the four benchmarks under 30-run execution. Tasks with $p \ge 0.05$ are considered non-significant. For tasks with $p < 0.05$, we further categorize the effect sizes using Cliff's delta ($\delta$): negligible ($\delta < 0.147$), small ($0.147 \le \delta < 0.33$), medium ($0.33 \le \delta < 0.474$), and large ($\delta \ge 0.474$).}
\label{tab:rq1}
\setlength{\tabcolsep}{4pt}
\renewcommand{\arraystretch}{1.05}

\begin{tabular}{l c c cccc}
\toprule
\multirow{2}{*}{\textbf{Benchmark}} &
\multirow{2}{*}{\textbf{\# Tasks}} &
\multirow{2}{*}{\makecell{\textbf{$p \ge 0.05$} \\ \textbf{(\# \%)}}} &
\multicolumn{4}{c}{\textbf{Effect Size $\delta$ ($p < 0.05$)}} \\
\cmidrule(lr){4-7}
&
&
&
\makecell{\textbf{Neg.}} &
\makecell{\textbf{Small}} &
\makecell{\textbf{Medium}} &
\makecell{\textbf{Large}} \\
\midrule

EffiBench & 1,000 & 944 (94.4\%) & 0 & 0 & 0 & 56 (5.6\%) \\

Enamel    &   164 & 153 (93.29\%) & 0 & 1 (0.61\%) & 1 (0.61\%) & 9 (5.49\%) \\

EvalPerf  &   118 & 102 (86.44\%) & 0 & 0 & 0 & 16 (13.56\%) \\

Mercury   &   256 & 245 (95.7\%)  & 0 & 0 & 0 & 11 (4.3\%) \\

\midrule
\textbf{Total} & \textbf{1,538} & 1,444 (93.89\%) & 0 & 1 (0.07\%) & 1 (0.07\%) & 92 (5.98\%) \\
\bottomrule
\end{tabular}
\end{table}

\phead{Results.}
Table~\ref{tab:rq1} summarizes the results across all four benchmarks. Across all 1,538 tasks, only 94 (6.11\%) tasks' performant implementations are significantly faster than the corresponding canonical solutions, while the remaining 1,444 (93.89\%) tasks show no statistically significant difference. This pattern is consistent across EffiBench, Enamel, and Mercury where over 93\% of tasks in each benchmark show no statistically significant difference. Among the tasks with $p < 0.05$, most exhibit large effect sizes (92 cases, 5.98\%), with only 1 small (0.07\%) and 1 medium (0.07\%) case, and no negligible effects observed. Three out of four benchmarks (i.e., EffiBench, Enamel, and Mercury) have fewer than 10\% tasks where the performance differences are statistically significant. EvalPerf has the highest rate of detectable performance improvement (13.56\%). This may be attributed to the fact that EvalPerf was specifically designed to select performance-challenging tasks and generate computationally expensive inputs. However, a large portion of EvalPerf's tasks (86.44\%) still remain non-statistically significant. 

We also compare the average mean runtime differences across tasks for each benchmark and apply the Wilcoxon signed-rank test to examine statistical differences. Across all benchmarks, over 93\% of tasks do not exhibit statistically significant differences between performant implementations and canonical solutions, with p-values often close to 1. When restricting the analysis to the small subset of statistically significant tasks ($p < 0.05$), the performant implementations are on average faster than the canonical solutions, with mean improvements ranging from +6.6\% to +27.1\% and p-values below $0.001$ with medium to large effect sizes.

\rqboxc{{\bf RQ1-Takeaway.}
Under repeated execution (30 runs) and statistical testing, over 93\% of tasks show no statistically significant performance differences between performant implementations and canonical solutions. Only a small fraction (6.11\%) exhibit measurable improvements, indicating that most benchmark-provided performant implementations do not yield actual performance gains under existing test suites.}
\subsection{RQ2: Do Code Changes in Performant Implementations Have Real Impact?}

\phead{Motivation.}
RQ1 establishes that 93.89\% of benchmark tasks' performant implementations are not statistically significantly faster than the corresponding canonical solutions. This raises a fundamental question: do benchmark-provided performant implementations, change throughout implementations actually involve code changes with performance impact?

\phead{Approach.}
To investigate this research question, we first performed open coding on a subset sampled from the 1,444 non-significant tasks identified RQ1 to identify the code change patterns between the canonical solution and the corresponding performant implementation. We used stratified sampling~\citep{neyman1992two, baltesR22} to preserve representativeness across benchmarks and sampled 308 tasks. The sample size is statistically significant, with a 95\% confidence level with a 5\% margin of error. 
Before performing open coding, two authors met and reviewed several pairs of the canonical solution and the performant implementation together. At this stage, we found that code change patterns could be grouped into three categories: (1) \textit{changes without performance impact}, (2) \textit{changes with performance impact}, and (3) \textit{no changes} (i.e., the canonical solution and the performant implementation are identical).

For each sampled task, two authors then independently compared the canonical solutions and the performant implementations. They were asked to classify each task into one of the three aforementioned categories. In addition, they were tasked with summarizing and documenting the code change patterns in greater detail. These two authors held multiple meetings to develop a codebook and resolve labeling inconsistencies. Eventually, in addition to the group of \textit{no changes}, two themes were identified for \textit{changes without performance impact}, and two themes were identified for \textit{changes with performance impact}. The final Cohen's Kappa~\citep{cohen1960coefficient} is 0.93, indicating an almost perfect agreement~\citep{landis1977measurement}. The whole labeling process took around 1200 person hours.

\iuhead{LLM-as-a-Judge.}
After establishing the root-cause categories through manual analysis, we use LLM-as-a-Judge as a secondary validation step.
Its role is not to replace the manual findings, but rather to \emph{confirm} whether the discovered categories can be reliably recognized by strong language models and to extend the analysis to the remaining 1,136 non-significant tasks across the four benchmarks. 

Recent studies~\citep{ahmedDTP25,lillmjudge2024,zhangLWSMS024} suggest that LLM-based evaluators can achieve agreement levels comparable to human annotators in structured classification tasks.
Motivated by this, we design a Chain-of-Thought prompting strategy that guides the model through a structured comparison of canonical solutions and benchmark-provided performant implementations.
For each task, the model is asked to reason about the differences (if any) between the two implementations in terms of algorithmic strategy, data structure choices, implementation-level optimizations, and input-dependent behavior, and then group the task into one of the themes we identified through manual analysis~\citep{artifact}.

We evaluated two judge models, \textit{gpt-4o-2024-08-06}~\citep{openai_chatgpt_4o_2024_08_06} and \textit{deepseek-v3.1}~\citep{deepseek_chat_2025}, with the temperature set to zero in all runs to minimize randomness and improve consistency.
To assess the reliability of LLM-as-a-Judge, we first applied LLMs to classify the 308 manually labeled tasks. Then we computed Cohen's Kappa~\citep{cohen1960coefficient} to measure the inter-rater agreement between the LLM and the human labelers. The scores are 0.75 and 0.72 with GPT-4o-2024-08-06 and DeepSeek-v3.1, respectively, indicating substantial agreement~\citep{landis1977measurement}. These results indicate that LLM-as-a-Judge can reliably reproduce manual labeling. We then applied LLM-as-a-Judge to label the remaining 1,136 tasks.

\begin{table}[t]
\centering
\caption{Code change patterns between the canonical solutions and benchmark-provided performant implementations.}
\label{tab:root-causes-perf-diff}

\setlength{\tabcolsep}{2pt}
\renewcommand{\arraystretch}{1.15}
\begin{tabular}{@{}>{\raggedright\arraybackslash}m{2.5cm} >{\raggedright\arraybackslash}m{2.8cm} c c c@{}}
\toprule

\multirow{2}{*}{\textbf{Change pattern}} &
\multirow{2}{*}{\textbf{Subcategory}} &
\multirow{2}{*}{\textbf{Manual (308)}} &
\multicolumn{2}{c}{\textbf{LLM-as-a-Judge (1,136)}} \\

\cmidrule(l){4-5}

& & &
\makecell{\textbf{DeepSeek-}\\\textbf{v3.1}} &
\makecell{\textbf{GPT-4o-}\\\textbf{2024-08-06}} \\

\midrule

\multicolumn{2}{@{}l}{\textbf{\textit{Changes without performance impact}}} &
\textbf{84 (27.27\%)} & \textbf{204 (17.96\%)} & \textbf{218 (19.19\%)} \\
\midrule
\multicolumn{2}{@{}l}{Built-in function substitution} & 4 (1.3\%) & 0 (0\%) & 1 (0.09\%) \\
\multicolumn{2}{@{}l}{Refactoring} & 80 (25.97\%) & 204 (17.96\%) & 217 (19.1\%) \\

\midrule

\multicolumn{2}{@{}l}{\textbf{\textit{Changes with performance impact}}} &
\textbf{209 (67.86\%)} & \textbf{864 (76.06\%)} & \textbf{840 (73.94\%)} \\
\midrule
\multirow{3}{2.4cm}{Data structure replacement} & Array to Hash Map / Set & 24 (7.79\%) & 39 (3.43\%) & 56 (4.93\%) \\
& \cellcolor{lightgray} Array to Heap / Priority Queue & \cellcolor{lightgray} 2 (0.65\%) & \cellcolor{lightgray} 6 (0.53\%)  & \cellcolor{lightgray} 8 (0.7\%) \\
\midrule
\multirow{12}{2.4cm}{Algorithm strategy change} & Sorting-Based Optimization & 22 (7.14\%) & 148 (13.03\%) & 132 (11.62\%) \\
& \cellcolor{lightgray}Dynamic Programming & \cellcolor{lightgray} 34 (11.04\%) & \cellcolor{lightgray} 121 (10.65\%) & \cellcolor{lightgray} 99 (8.71\%) \\
& Graph Traversal (BFS/DFS) & 26 (8.44\%) & 92 (8.1\%) & 119 (10.48\%) \\
& \cellcolor{lightgray}Binary Search & \cellcolor{lightgray} 11 (3.57\%) & \cellcolor{lightgray} 36 (3.17\%) & \cellcolor{lightgray} 43 (3.79\%) \\
& Mathematical Optimization & 19 (6.17\%) & 105 (9.24\%) & 95 (8.36\%) \\
& \cellcolor{lightgray}Prefix Sum & \cellcolor{lightgray} 4 (1.3\%) & \cellcolor{lightgray} 37 (3.26\%) & \cellcolor{lightgray} 29 (2.55\%) \\
& Sliding Window / Two Pointers & 12 (3.9\%) & 99 (8.71\%) & 73 (6.43\%) \\
& \cellcolor{lightgray}Control-Flow Restructuring & \cellcolor{lightgray} 22 (7.14\%) & \cellcolor{lightgray} 23 (2.02\%) & \cellcolor{lightgray} 34 (3\%) \\
& Slice-Based Vectorization & 12 (3.9\%) & 28 (2.46\%) & 42 (3.7\%) \\
& \cellcolor{lightgray} Bit Manipulation & \cellcolor{lightgray} 11 (3.57\%) & \cellcolor{lightgray} 54 (4.75\%) & \cellcolor{lightgray} 49 (4.31\%) \\
& Stack-Based Approach & 4 (1.3\%) & 40 (3.52\%) & 28 (2.46\%) \\
& \cellcolor{lightgray}Other & \cellcolor{lightgray} 6 (1.95\%) & \cellcolor{lightgray} 36 (3.17\%) & \cellcolor{lightgray} 33 (2.9\%) \\

\midrule

\textbf{No changes} & & \textbf{15 (4.87\%)} & \textbf{68 (6\%)} & \textbf{78 (6.9\%)} \\

\bottomrule
\end{tabular}
\end{table}

\phead{Results.}
Table~\ref{tab:root-causes-perf-diff} presents the code change patterns we identified through our manual analysis and the distribution across the dataset. We elaborate on four code change patterns we identified. 

\iuhead{Changes without performance impact.} 84 out of the 308 tasks were classified with changes that do not likely to have impact on runtime performance. We identified two specific patterns:

\begin{itemize}[leftmargin=*]

    \item \textbf{Built-in function substitution} replaces a manual implementation or library call with an equivalent built-in function that provides the same functionality without changing performance behavior. For example, replacing the manual loops with built-in \texttt{sum}, \texttt{max}, \texttt{count} for arrays.

    4 out of the 308 tasks belong to this group. In these cases, both program semantics and asymptotic complexity are preserved, the runtime performance should not have significant differences.

    \item \textbf{Refactoring} include changes that improve readability or organization (e.g., renaming, reordering conditions, and restructuring control flow) without modifying semantics or performance behavior. For example, renaming variables such as \textit{balance} to \textit{b} or \textit{max\_reachable} to \textit{mx}, or rewriting loops from \textit{for i in range(len(nums))} to \textit{for x in nums}.

    We identified 80 out of the 308 tasks involves such refactoring. These changes have no performance impact, as complexity remains unchanged. 
    
\end{itemize}

\iuhead{Changes with performance impact.} 209 out of the 308 tasks involves changes that may have impact on runtime performance. Specifically, the following two patterns were found:

\begin{itemize}[leftmargin=*]

    \item \textbf{Algorithm strategy change} includes changes that replace one algorithmic approach with another that significantly changes execution behavior or reduces computational complexity.

        
     For example, a brute-force enumeration may be replaced with a sorting-based approach that avoids repeatedly exploring the same search space. In our manual analysis, 183 out of the 308 tasks (59.42\%) include algorithm strategy changes. The detailed subcategories in Table~\ref{tab:root-causes-perf-diff} show that these changes cover a range of optimization strategies. Dynamic programming and graph traversal are the two most common subcategories, appearing in 34 tasks (11.04\%) and 26 tasks (8.44\%), respectively. The remaining cases are distributed across sorting-based optimization and control-flow restructuring, each with 22 tasks (7.14\%); mathematical optimization, with 19 tasks (6.17\%); sliding window or two-pointer techniques and slice-based vectorization, each with 12 tasks (3.9\%); binary search and bit manipulation, each with 11 tasks (3.57\%); and prefix sums and stack-based approaches, each with 4 tasks (1.3\%). The Other subcategory contains 6 tasks (1.95\%) whose algorithmic changes do not belong to any of the listed categories. Although these changes may have performance impact, the benchmark-provided test cases with inputs are often not effective enough to distinguish the runtime differences between the canonical and performant implementations.

    \item \textbf{Data structure replacement} includes changes where the performant implementation uses a different data structure to reduce the cost of repeated operations, such as replacing linear scans over arrays with hash-based lookup using maps or sets, or using heaps or priority queues to manage ordered elements more efficiently.

    26 out of the 308 manually analyzed tasks (8.44\%) include data structure replacements. Among these tasks, replacing arrays with hash maps or sets is substantially more common than replacing arrays with heaps or priority queues: we observe 24 array-to-hash-map/set replacements (7.79\%) and 2 array-to-heap/priority-queue replacements (0.65\%). However, their runtime benefits can remain hidden when the benchmark-provided test cases with inputs are not large or complex enough for these operations to dominate execution time.

\end{itemize}

Surprisingly, we also found 15 tasks where the canonical solution and the performant implementation are identical. These results indicate that further data cleaning could be essential for existing benchmarks.

Our findings are further supported by the expanded analysis using LLM-as-a-Judge on all remaining 1,136 non-significant tasks.  17.96\% and 19.19\% of the tasks are classified as changes without performance impact by DeepSeek-v3.1 and GPT-4o-2024-08-06, respectively. Refactoring remains the most frequent change pattern that is unlikely to have a performance impact, identified in 17.96\% of tasks by DeepSeek-v3.1 and 19.19\% by GPT-4o-2024-08-06, confirming that this pattern is pervasive across the full dataset. In terms of tasks involving changes that may have performance impact, 72.1\% and 68.31\% of the tasks are identified as involving algorithm strategy changes by DeepSeek-v3.1 and GPT-4o-2024-08-06, respectively. 3.96\% and 5.63\% of the tasks are identified as involving data structure replacements, respectively. These results indicate that the performance implementations for the majority of tasks in existing benchmarks include changes with potential performance impact.  However, these changes do not result in statistically significant runtime differences under the current evaluation settings

\rqboxc{{\bf RQ2-Takeaway 1.}
Both manual analysis and LLM-as-a-Judge analysis confirm that about 21.47\% of the tasks in existing benchmarks do not include meaningful performance changes compared to the canonical solutions, while about 72.62\% involve potential performance improvements that do not result in statistically significant runtime differences under the current evaluation settings.}

{
\setlength\fboxsep{0pt}

\begin{samepage}
\begin{center}
\begin{minipage}[t]{0.48\linewidth}
\noindent\makebox[\linewidth][l]{\textbf{Canonical solution} \textcolor{gray}{$O(n^3)$}}
\begin{lstlisting}[language=Python, escapechar=!, numberstyle=\tiny\color{lightgray}, basicstyle=\scriptsize\ttfamily]
def triples_sum_to_zero(l):
!{\colorbox{yellow!30}{    \codeoperation{for} i \codeoperation{in} \codeoperation{range}(\codeoperation{len}(l)):}}!
!{\colorbox{yellow!30}{        \codeoperation{for} j \codeoperation{in} \codeoperation{range}(i + 1, \codeoperation{len}(l)):}}!
!{\colorbox{yellow!30}{            \codeoperation{for} k \codeoperation{in} \codeoperation{range}(j + 1, \codeoperation{len}(l)):}}!
                if l[i] + l[j] + l[k] == 0:
                    return True
    return False
\end{lstlisting}
\end{minipage}
\hfill
\begin{minipage}[t]{0.48\linewidth}
\noindent\makebox[\linewidth][l]{\textbf{Performant implementation} \textcolor{gray}{$O(n^2)$}}
\begin{lstlisting}[language=Python, escapechar=!, numberstyle=\tiny\color{lightgray}, basicstyle=\scriptsize\ttfamily]
def triples_sum_to_zero(l):
     !{\colorbox{yellow!30}{\codeoperation{for} i, x \codeoperation{in} \codeoperation{enumerate}(l[:-2]):}}!
        buf = set()
        !{\colorbox{yellow!30}{\codeoperation{for} y \codeoperation{in} l[i + 1:]:}}!
            if y in buf:
                return True
            buf.add(-x - y)
    return False
\end{lstlisting}
\end{minipage}
\captionsetup{type=lstlisting,hypcap=false}
\caption{Algorithm strategy changes improve time complexity from $O(n^3)$ to $O(n^2)$}
\label{exp:algorithm_change}
\end{center}
\end{samepage}
}

To understand why current evaluation settings cannot reliably expose performance differences between performant implementations and canonical solutions, we reviewed the corresponding test suites for these tasks. We found that the benchmark-provided test suites are often too small. For instance, the performant implementation of Task 40 in the Enamel benchmark includes an algorithm strategy change (Example~\ref{exp:algorithm_change}). A set-based approach ($O(n^2)$) is used in the performant implementation instead of a triple-loop approach ($O(n^3)$) in the canonical solution. However, the length of the longest input array in the benchmark-provided test suite is 6, which is insufficient to distinguish the performance differences between the two implementations.

{
\setlength\fboxsep{0pt}

\begin{samepage}
\begin{center}
\begin{minipage}[t]{0.48\linewidth}
\noindent\makebox[\linewidth][l]{\textbf{Canonical solution} \textcolor{gray}{$O(n^2)$}}
\begin{lstlisting}[language=Python, escapechar=!, numberstyle=\tiny\color{lightgray}, basicstyle=\scriptsize\ttfamily]
def frequency_lists(list):
    list = [item for sublist in list
            for item in sublist]
    return {
        x: !{\colorbox{yellow!30}{list.count(x)}}!
        for x in list
    }
\end{lstlisting}
\end{minipage}
\hfill
\begin{minipage}[t]{0.48\linewidth}
\noindent\makebox[\linewidth][l]{\textbf{Performant implementation} \textcolor{gray}{$O(n)$}}
\begin{lstlisting}[language=Python, escapechar=!, numberstyle=\tiny\color{lightgray}, basicstyle=\scriptsize\ttfamily]
def frequency_lists(nested_lists):
    flat_list = list(chain(*nested_lists))
    return dict(
        !{\colorbox{yellow!30}{Counter(flat\_list)}}!
    )
\end{lstlisting}
\end{minipage}
\captionsetup{type=lstlisting,hypcap=false}
\caption{Replacing inefficient \texttt{list.count} with \texttt{Counter}}
\label{exp:3}
\end{center}
\end{samepage}
}

The performant implementation of Mbpp/97 in the EvalPerf benchmark demonstrates a data structure change (Example~\ref{exp:3}). The canonical solution repeatedly counts element frequencies using \texttt{list.count}, resulting in quadratic time complexity ($O(n^2)$). In contrast, the performant implementation replaces this approach with a hash-based aggregation using \texttt{Counter}, reducing the complexity to linear time ($O(n)$).
However, the benchmark-provided test suite contains only small nested lists, which are insufficient to expose the performance gap between the two implementations.

\rqboxc{{\bf RQ2-Takeaway 2.}
A large portion (more than 60\%) of the tasks’ performant implementations introduce substantive algorithmic or data structure changes that should yield runtime gains. However, these changes do not result in statistically significant runtime differences because the benchmark-provided test suites do not include input sizes at which such changes affect execution time. These findings indicate that the test suites of existing benchmarks should be improved to cover input sizes that make the expected runtime differences measurable}

\subsection{RQ3: Can Multi-Agent-Based Test Generation Reveal Hidden Performance Differences?}

\phead{Motivation.}
In RQ2, we found that the test suites of existing benchmarks may be further improved. In this RQ, we investigate whether and to what extent a multi-agent-based approach can enhance the test suites to better detect such performance gains.

\begin{figure}[htbp]
    \centering
    \makebox[\textwidth][c]{{\includegraphics[width=1\textwidth]{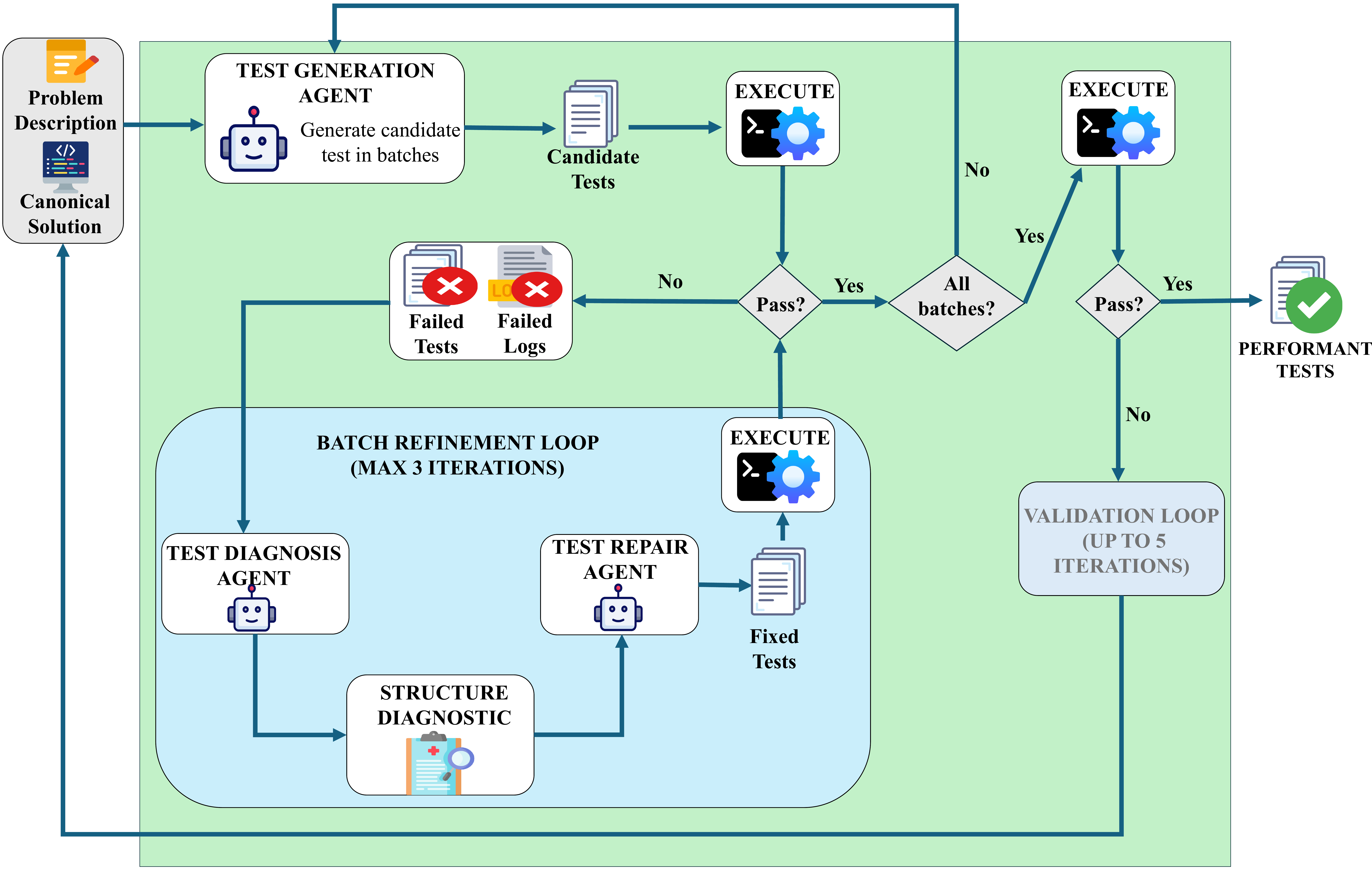}}}
    \caption{Overview of our multi-agent framework for generating and repairing performance-oriented test suites.}
    \label{agent_framework}
\end{figure}

\phead{Approach.} We develop a multi-agent framework to automatically generate performance-oriented test suites, as illustrated in Figure~\ref{agent_framework}. The framework contains three agents: The Test Generation Agent, The Diagnosis Agent, and The Repair Agent. The \textit{Test Generation Agent} receives the problem description and the code under test, then creates deterministic tests that preserve the intended behavior while using inputs that can expose runtime differences. The \textit{Test Diagnosis Agent} is used when generated tests fail. Its role is to reason over the error log and convert the full traceback into a compact structured root-cause diagnostic, including the error type and message, the root cause of the failure, and the relevant line information. This diagnostic reduces the log text included in the LLM prompt for the Test Repair Agent and provides clearer evidence
for fixing the failed tests. The \textit{Test Repair Agent} then uses the code, failed-test identifiers, and diagnostics to repair the failed tests.

Both DeepSeek-V3.1~\citep{deepseek_chat_2025} and GPT-4o-2024-08-06~\citep{openai_chatgpt_4o_2024_08_06} have a context window of 128,000 tokens. Therefore, we generate ten tests for each task in two batches of five rather than in a single request. This design bounds the amount of generated code, failure logs, and repair information included in each prompt while preserving sufficient space for the model's response.

\begin{figure}[t]
    \centering
    \begin{tcolorbox}[colback=blue!6!white, colframe=blue!55!black, title=\textbf{Prompt Template: Test Generation (Simplified)}, sharp corners, fontupper=\scriptsize]
\textbf{Task:} Generate exactly $y$ standalone performance-oriented test functions for the current batch.\\
\textbf{Instructions:}\\
1. Read the problem description and the code.\\
2. Create deterministic inputs that preserve the intended behavior but make inefficient code run longer.\\
3. Reason about the expected correct behavior independently.\\
4. Avoid timing assertions, I/O, randomness inside loops, external libraries, comments, and explanations.\\
\textbf{Output:} JSON with a \texttt{testcase} field containing the generated Python test functions.
\end{tcolorbox}

    \begin{tcolorbox}[colback=green!7!white, colframe=green!45!black, title=\textbf{Prompt Template: Test Diagnosis (Simplified)}, sharp corners, fontupper=\scriptsize]
\textbf{Task:} Analyze the failure of a generated test.\\
\textbf{Instructions:}\\
1. Read the failed test function, the code under test, and the error log.\\
2. Identify the error type and error message from the log.\\
3. Identify the root cause in the generated test using the evidence from the failure.\\
4. Reason about the traceback to identify the failing line number and line content from the test function.\\
5. Do not generate code and do not output the full traceback.\\
\textbf{Output:} JSON with \texttt{error}, \texttt{root\_cause}, \texttt{line\_number}, and \texttt{line\_content}.
\end{tcolorbox}

    \begin{tcolorbox}[colback=orange!9!white, colframe=orange!70!black, title=\textbf{Prompt Template: Test Repair (Simplified)}, sharp corners, fontupper=\scriptsize]
\textbf{Task:} Repair or regenerate only the failed generated tests.\\
\textbf{Instructions:}\\
1. Read the code under test, failed tests, and structured diagnostics.\\
2. Infer the intended behavior from the code and use the diagnostics to locate each test problem.\\
3. Fix invalid expected values, invalid inputs, nondeterminism, or timeout-prone inputs.\\
4. Preserve the original failed test names.\\
5. Do not modify the implementation code; avoid classes, external libraries, comments, explanations, stack traces, and large diffs.\\
\textbf{Output:} JSON with a \texttt{testcase} field containing the repaired Python test functions.
\end{tcolorbox}
    \caption{Simplified prompt templates used by the Test Generation, Test Diagnosis, and Test Repair Agents.}
    \label{fig:agent_prompts}
\end{figure}

For each batch, the Test Generation Agent produces five candidate tests, which are then executed against the canonical implementation. If some tests fail, the framework sends only the failing tests from that batch, the code under test, and the corresponding error logs to the Test Diagnosis Agent. This agent analyzes each failure to identify the error type and message, determine its root cause, and locate the relevant failing line in the generated test. It then converts this information into a compact structured diagnostic. The Test Repair Agent uses these diagnostics to correct the expected outputs, inputs, or input sizes of the failed tests. Restricting diagnosis and repair to the failed tests further controls prompt length and avoids repeatedly including the full generated suite. After repair, the batch is re-executed. This diagnose-and-repair loop is repeated for up to three refinement iterations per batch and stops early when all five tests pass.

After both batches are processed, the framework executes the combined suite of ten generated tests. If any test still fails, it repeats the batch-level generation and refinement process for only the failing tests, again providing the LLM with summarized diagnostics rather than the complete failure history. This full-suite validation loop is allowed up to five iterations and stops when all generated tests pass or the iteration limit is reached. If the iteration limit is reached while some tests still fail, the framework filters out the failing tests and keeps only those that pass validation. The final output for each task is therefore a validated set of executable, deterministic, performance-oriented tests that can be used to re-measure the canonical and performant implementations under the same statistical procedure as RQ1. Our simplified prompt templates can be found in Fig.~\ref{fig:agent_prompts}

We evaluate our framework against COFFE~\citep{coffee25}, a SOTA LLM-based performance test generation for function-level Python programs in two stages. First, we apply both approaches to the 209 tasks from the manual analysis that were confirmed to contain code changes with performance impact but yielded no significant results under the benchmark-provided test suites. This stage directly measures how many previously hidden performance gains each approach can reveal through its generated performance-oriented tests. We then
extend the evaluation to the remaining 1,136 non-significant tasks to assess their effectiveness across the dataset.

\begin{table}[htbp]
\centering
\caption{Two-stage evaluation of generated test suites on tasks that were non-significant under the benchmark-provided test suites, broken down by approach and benchmark. Stage 1 evaluates the 209 manually verified tasks whose code changes were confirmed to have performance impact. Stage 2 evaluates the remaining 1,136 non-significant tasks. We report the number of tasks that become statistically significant ($p<0.05$) under each generated test suite. All detected significant cases correspond to large effect sizes ($\delta \ge 0.474$).}
\label{tab:rq3}

\renewcommand{\arraystretch}{1.05}
\setlength{\tabcolsep}{1.5pt}
\footnotesize
\begin{tabular}{l l c c c c c}
\toprule
\textbf{Stage} & \textbf{Benchmark} & \textbf{\#Tasks} &
\multicolumn{2}{c}{\textbf{Our approach}} &
\multicolumn{2}{c}{\textbf{COFFE}} \\
\cmidrule(lr){4-5}
\cmidrule(l){6-7}
& & &
\makecell{\textbf{DeepSeek}} &
\makecell{\textbf{GPT-4o}} &
\makecell{\textbf{DeepSeek}} &
\makecell{\textbf{GPT-4o}} \\
\midrule

\multirow{5}{*}{\makecell[l]{Manual\\Verified}}
& EffiBench & 138 & 51 (36.96\%) & 63 (45.65\%) & 3 (2.17\%) & 0 (0\%) \\
& Enamel    & 18 & 14 (77.78\%) & 13 (72.22\%) & 0 (0\%) & 0 (0\%) \\
& EvalPerf  & 19 & 14 (73.68\%) & 12 (63.16\%) & 0 (0\%) & 1 (5.26\%) \\
& Mercury   & 34 & 15 (44.12\%) & 14 (41.18\%) & 1 (2.94\%) & 0 (0\%) \\
& \textbf{Total} & \textbf{209} & \textbf{94 (44.98\%)} & \textbf{102 (48.80\%)} & \textbf{4 (1.91\%)} & \textbf{1 (0.48\%)} \\
\midrule

\multirow{5}{*}{\makecell[l]{Remaining}}
& EffiBench & 744 & 108 (14.52\%) & 115 (15.46\%) & 15 (2.01\%) & 0 (0\%) \\
& Enamel    & 120 & 72 (60\%) & 69 (57.5\%) & 0 (0\%) & 0 (0\%) \\
& EvalPerf  & 78 & 23 (29.49\%) & 26 (33.33\%) & 0 (0\%) & 5 (6.41\%) \\
& Mercury   & 194 & 26 (13.40\%) & 30 (15.46\%) & 8 (4.12\%) & 0 (0\%) \\
& \textbf{Total} & \textbf{1,136} & \textbf{229 (20.16\%)} & \textbf{240 (21.13\%)} & \textbf{23 (2.02\%)} & \textbf{5 (0.44\%)} \\
\midrule

\textbf{Overall} & \textbf{Total} & \textbf{1,345} & \textbf{323 (24.01\%)} & \textbf{342 (25.43\%)} & \textbf{27 (2\%)} & \textbf{6 (0.45\%)} \\
\bottomrule
\end{tabular}
\end{table}

\phead{Results.}
Table~\ref{tab:rq3} shows that our framework consistently enables more tasks to exhibit statistically significant performance differences than COFFE. On the 209 manually verified tasks, our DeepSeek-v3.1 and GPT-4o test suites yield statistically significant performance differences for 94 (44.98\%) and 102 (48.80\%) tasks, respectively. Under the same models, COFFE achieves this outcome for only 4 (1.91\%) and 1 (0.48\%) tasks. Thus, for the subset in which the code changes are known to affect performance, our framework reveals substantially more of the performance differences that were not observable with the benchmark-provided test suites.

The same pattern holds for the remaining 1,136 tasks. Our framework yields statistically significant performance differences for 229 (20.16\%) tasks with DeepSeek-v3.1 and 240 (21.13\%) with GPT-4o, compared with 23 (2.02\%) and 5 (0.44\%) for COFFE. Overall, across all 1,345 tasks, our framework achieves statistically significant results for 323 (24.01\%) and 342 (25.43\%) tasks, whereas COFFE does so for only 27 (2.00\%) and 6 (0.45\%), respectively. All statistically significant cases have large effect sizes ($\delta \ge 0.474$). These results indicate that our multi-agent generation, diagnosis, and repair process produces performance-oriented test suites that expose performance differences substantially more effectively than COFFE in our Python function-level setting.

{
\setlength\fboxsep{0pt}

\begin{samepage}
\begin{center}
\begin{minipage}[t]{0.48\linewidth}
\noindent\makebox[\linewidth][l]{\textbf{Canonical solution}}
\begin{lstlisting}[language=Python, escapechar=!, numberstyle=\tiny\color{lightgray}, basicstyle=\scriptsize\ttfamily]
def lengthOfLongestSubstring(s):
    !{\colorbox{yellow!30}{ss, i, ans = set(), 0, 0}}!
    for j, c in enumerate(s):
        !{\colorbox{yellow!30}{\codeoperation{while} c \codeoperation{in} ss:}}!
            ss.remove(s[i]); i += 1
        ss.add(c)
        ans = max(ans, j - i + 1)
    return ans
\end{lstlisting}
\end{minipage}
\hfill
\begin{minipage}[t]{0.48\linewidth}
\noindent\makebox[\linewidth][l]{\textbf{Performant implementation}}
\begin{lstlisting}[language=Python, escapechar=!, numberstyle=\tiny\color{lightgray}, basicstyle=\scriptsize\ttfamily]
def lengthOfLongestSubstring(s):
    !{\colorbox{yellow!30}{idx, start, ans = \{\}, 0, 0}}!
    for i, char in enumerate(s):
        !{\colorbox{yellow!30}{\codeoperation{if} char \codeoperation{in} idx:}}!
            !{\colorbox{yellow!30}{\codeoperation{if} idx[char] >= start:}}!
                start = idx[char] + 1
        idx[char] = i
        ans = max(ans, i - start + 1)
    return ans
\end{lstlisting}
\end{minipage}
\begin{lstlisting}[language=Python, escapechar=!, numberstyle=\tiny\color{lightgray}, basicstyle=\scriptsize\ttfamily]
# Original tests - small inputs
assert lengthOfLongestSubstring('uIxsZwqW2u') == 9
# Generated tests - large inputs
assert lengthOfLongestSubstring('ab' * 500000) == 2
assert lengthOfLongestSubstring('a' * 1000000) == 1
\end{lstlisting}
\captionsetup{type=lstlisting,hypcap=false}
\caption{Our generated tests effectively expose performance differences with larger inputs}
\label{exp:task3}
\end{center}
\end{samepage}
}

Task 3 of EffiBench illustrates how our generated tests can reveal a performance difference that is not statistically significant under the benchmark-provided tests. The performant implementation uses index jumps instead of incremental window shrinking (Example~\ref{exp:task3}). However, the benchmark inputs are too short (the maximum input-string length is 10), causing the two implementations to exhibit similar execution times. In contrast, our generated tests use much larger inputs (e.g., a string of length 500,000), making the performance advantage of index jumps observable.

A key limitation of our approach arises when handling programs with multiple nested functions, where the agent may misidentify the target function for testing. As shown in Example~\ref{exp:example5}, the canonical solution defines \texttt{sum\_Of\_product} as the entry-point, which calls \texttt{binomial\_Coeff}. However, the agent incorrectly targets \texttt{binomial\_Coeff} during test generation, as it appears to contain the core logic.
This leads to a mismatch when evaluating the performant implementation, where the computation is implemented directly within \texttt{sum\_Of\_product} and \texttt{binomial\_Coeff} no longer exists. As a result, the generated tests become incompatible, preventing meaningful performance comparison.
This issue highlights the agent's limitations in handling functional interfaces and suggests the need for improved function-level reasoning to correctly identify and target the intended entry-point.

\begin{figure}[htbp]
\centering
\setlength\fboxsep{0pt}
\begin{minipage}[t]{0.48\linewidth}
\noindent\makebox[\linewidth][l]{\textbf{Canonical solution} \textcolor{gray}{$O(n^2)$}}
\begin{lstlisting}[language=Python, escapechar=!, numberstyle=\tiny\color{lightgray}, basicstyle=\scriptsize\ttfamily]
def binomial_Coeff(n, k):
    C = [0] * (k + 1)
    C[0] = 1
    for i in range(1, n + 1):
        for j in range(min(i, k), 0, -1):
            C[j] = C[j] + C[j - 1]
    return C[k]

!{\colorbox{yellow!30}{\codeoperation{def} sum\_Of\_product(n):}}!
    return binomial_Coeff(2*n, n-1)
\end{lstlisting}
\end{minipage}
\hfill
\begin{minipage}[t]{0.48\linewidth}
\noindent\makebox[\linewidth][l]{\textbf{Performant implementation} \textcolor{gray}{$O(n)$}}
\begin{lstlisting}[language=Python, escapechar=!, numberstyle=\tiny\color{lightgray}, basicstyle=\scriptsize\ttfamily]
!{\colorbox{yellow!30}{\codeoperation{def} sum\_Of\_product(n):}}!
    binomial, binomial[0] = [0] * (n + 1), 1
    total = 0
    for i in range(1, n + 1):
        binomial[i] = binomial[i-1] * (n+1-i) // i
        total += binomial[i-1] * binomial[i]
    return total
\end{lstlisting}
\end{minipage}
\begin{lstlisting}[language=Python, escapechar=!, numberstyle=\tiny\color{lightgray}, basicstyle=\scriptsize\ttfamily]
# Original test cases
assert !{\colorbox{yellow!30}{sum\_Of\_product(3)}}! == 15
# Generated test cases - wrong function
expected = 8965199470901314966...0932000
assert !{\colorbox{yellow!30}{binomial\_Coeff(200, 99)}}! == expected
# Fixed generated tests
assert sum_Of_product(100) == expected
\end{lstlisting}
\captionsetup{type=lstlisting,hypcap=false}
\caption{Our generated tests fail to expose performance differences}
\label{exp:example5}
\end{figure}

\rqboxc{{\bf RQ3-Takeaway.}
Our generated tests uncover 44.98\% and 48.80\% of previously missed performance differences on the manually verified subset with performance impact, and 24.01\% and 25.43\% on the overall dataset, using DeepSeek-v3.1 and GPT-4o, respectively. These results demonstrate the effectiveness of our approach in generating performance-oriented test suites.}
\subsection{RQ4: How do performance-oriented test suites affect the benchmark results?}

\begin{table*}[p]
\centering
\makebox[0pt][c]{%
\rotatebox{90}{%
\begin{minipage}{\textheight}
\centering
\caption{Statistical analysis results across four benchmarks under repeated execution (30 runs) for three LLM-generated code models. For each task, we apply the Mann--Whitney U test. Tasks with $p \ge 0.05$ are considered not statistically significant. For tasks with $p < 0.05$, effect sizes are categorized using Cliff's delta ($\delta$): negligible ($\delta < 0.147$), small ($0.147 \le \delta < 0.33$), medium ($0.33 \le \delta < 0.474$), and large ($\delta \ge 0.474$).}
\label{tab:rq4}
\begingroup
\renewcommand{\arraystretch}{0.95}
\setlength{\tabcolsep}{1pt}
\small

\begin{tabular}{l l c c cccc c cccc c cccc}
\toprule
\textbf{Model} &
\textbf{Benchmark} &
\textbf{\#Tasks} &

\multicolumn{5}{c}{\textbf{Original Tests}} &
\multicolumn{5}{c}{\textbf{DeepSeek-v3.1 Tests}} &
\multicolumn{5}{c}{\textbf{GPT-4o Tests}} \\

\cmidrule(lr){4-8}
\cmidrule(lr){9-13}
\cmidrule(lr){14-18}

&
&
&
\makecell{\textbf{$p \ge 0.05$}} &
\multicolumn{4}{c}{\makecell{\textbf{Effect size} \\ ($p < 0.05$)}} &

\makecell{\textbf{$p \ge 0.05$}} &
\multicolumn{4}{c}{\makecell{\textbf{Effect size} \\ ($p < 0.05$)}} &

\makecell{\textbf{$p \ge 0.05$}} &
\multicolumn{4}{c}{\makecell{\textbf{Effect size} \\ ($p < 0.05$)}} \\

\cmidrule(lr){5-8}
\cmidrule(lr){10-13}
\cmidrule(lr){15-18}

&
&
&
&
\textbf{Neg.} & \textbf{Small} & \textbf{Med.} & \textbf{Large} &
&
\textbf{Neg.} & \textbf{Small} & \textbf{Med.} & \textbf{Large} &
&
\textbf{Neg.} & \textbf{Small} & \textbf{Med.} & \textbf{Large} \\

\midrule

\multirow{5}{*}{GPT-4o-mini}
& EffiBench & 1,000 & 947 & 0 & 0 & 0 & 53 & 777 & 0 & 0 & 2 & 221 & 772 & 0 & 0 & 1 & 227 \\
& Enamel    & 164  & 157 & 0 & 0 & 0 & 7 & 111 & 0 & 0 & 1 & 52 & 114 & 0 & 0 & 0 & 50 \\
& EvalPerf  & 118  & 108 & 0 & 0 & 0 & 10 & 81 & 0 & 1 & 0 & 36 & 80 & 0 & 0 & 0 & 38 \\
& Mercury   & 256  & 230 & 0 & 0 & 1 & 25 & 173 & 0 & 0 & 0 & 83 & 170 & 0 & 0 & 0 & 86 \\
& \textbf{Total} & \textbf{1,538} & 1,442 & 0 & 0 & 1 & 95 & 1,142 & 0 & 1 & 3 & 392 & 1,136 & 0 & 0 & 1 & 401 \\
\midrule

\multirow{5}{*}{Claude-Sonnet-4.5}
& EffiBench & 1,000 & 916 & 0 & 0 & 0 & 84 & 555 & 0 & 0 & 0 & 445 & 546 & 0 & 0 & 0 & 454 \\
& Enamel    & 164  & 154 & 0 & 0 & 0 & 10 & 95 & 0 & 0 & 0 & 69 & 92 & 0 & 0 & 0 & 72 \\
& EvalPerf  & 118  & 106 & 0 & 0 & 0 & 12 & 66 & 0 & 0 & 0 & 52 & 70 & 0 & 0 & 0 & 48 \\
& Mercury   & 256  & 212 & 0 & 0 & 0 & 44 & 142 & 0 & 0 & 0 & 114 & 132 & 0 & 0 & 0 & 124 \\
& \textbf{Total} & \textbf{1,538} & 1,388 & 0 & 0 & 0 & 150 & 858 & 0 & 0 & 0 & 680 & 840 & 0 & 0 & 0 & 698 \\
\midrule

\multirow{5}{*}{Gemini-2.5-Flash}
& EffiBench & 1,000 & 961 & 0 & 0 & 0 & 39 & 882 & 0 & 0 & 0 & 118 & 888 & 0 & 0 & 0 & 112 \\
& Enamel    & 164  & 154 & 0 & 0 & 0 & 10 & 108 & 0 & 0 & 0 & 56 & 111 & 0 & 0 & 0 & 53 \\
& EvalPerf  & 118  & 103 & 0 & 0 & 0 & 15 & 70 & 0 & 0 & 0 & 48 & 73 & 0 & 0 & 0 & 45 \\
& Mercury   & 256  & 240 & 0 & 0 & 0 & 16 & 204 & 0 & 0 & 0 & 52 & 204 & 0 & 0 & 0 & 52 \\
& \textbf{Total} & \textbf{1,538} & 1,458 & 0 & 0 & 0 & 80 & 1,264 & 0 & 0 & 0 & 274 & 1,276 & 0 & 0 & 0 & 262 \\

\bottomrule
\end{tabular}
\endgroup
\end{minipage}%
}%
}
\end{table*}

\phead{Motivation.}
In RQ3, we investigated whether a multi-agent approach can generate test suites under which more canonical and performant implementations exhibit statistically significant runtime differences. In this RQ, we examine how these performance-oriented test suites influence the evaluation results of LLM-generated code on the benchmarks.

\phead{Approach.}
To assess how stronger test suites affect benchmark evaluation results, we compare LLM generated implementations with the original implementations using both the benchmark provided test suites and the performance-oriented test suites introduced in RQ3. For each task, we consider performant implementations generated by three LLMs: \textit{GPT-4o-mini}~\citep{gpt_4o_mini}, \textit{Claude-Sonnet-4.5}~\citep{claude_sonnet_4_5}, and \textit{Gemini-2.5-Flash}~\citep{gemini_2_5_flash}, and evaluate them alongside canonical solutions under these test suites.

Following the statistical evaluation protocol used in RQ3, we run each implementation 30 times on each test suite and collect runtime measurements across runs. We then apply the same statistical testing method to determine whether the LLM generated implementation shows a statistically significant performance improvement over the original implementation. By comparing the results obtained from the benchmark tests and the performance-oriented tests across all tasks, we analyze how stronger test suites influence the measured efficiency improvements and the overall benchmark evaluation outcomes.

\phead{Results.}
Table~\ref{tab:rq4} summarizes the results across all four benchmarks for the three LLM-generated models under different test suites. Under the benchmark-provided test suites (Original), all three models show a similar pattern where most tasks do not exhibit statistically significant performance differences. This includes 1,442 (93.76\%) tasks for GPT-4o-mini, 1,388 (90.25\%) for Claude-Sonnet-4.5 and 1,458 (94.8\%) for Gemini-2.5-Flash. Among the significant cases ($p < 0.05$), almost all exhibit large effect sizes while negligible and small effects are nearly absent.

When evaluated with stronger test suites, the number of statistically distinguishable performance improvements increases substantially. Under DeepSeek-v3.1-generated tests, the number of significant cases increases to 396 (25.75\%), 680 (44.21\%), and 274 (17.82\%) for GPT-4o-mini, Claude-Sonnet-4.5, and Gemini-2.5-Flash, respectively. Under GPT-4o–generated tests, the number of significant cases increases to 402 (26.14\%), 698 (45.38\%), and 262 (17.03\%), respectively. Notably, Claude-Sonnet-4.5 shows a larger improvement compared to the other models. This may be attributed to the fact that Claude-Sonnet-4.5 usually has a longer and more complete reasoning process, which could enable it to generate more effective optimizations that are better exposed by stronger test suites. 
However, we also notice that many tasks still remain non-significant across all models, indicating the generation of performant implementations remain challenging for LLMs.

\rqboxc{{\bf RQ4-Takeaway.}
Modern LLMs are capable of generating performant implementations. Yet, their performance improvements could be ignored by the weak test suites provided by existing benchmarks. With our enhanced test suites generated in RQ3, an average of 22.19\% performance improvements can be further confirmed as statistically distinguishable.}

\section{Discussion and Future Work}
\label{sec:discussion}

In this section, we discuss the implications of our empirical findings on existing performance benchmarks and outline future research directions motivated by the limitations observed in our study.

\phead{Performance-Aware Benchmark Design.}
Our findings in RQ1 suggest that many tasks in existing efficiency benchmarks fail to reliably distinguish performance differences between canonical and performant implementations. Although these benchmarks aim to evaluate efficiency, most reported performance improvements are not statistically significant under repeated execution. This may be largely attributed to the fact that tasks of these benchmarks are adopted directly from datasets that prioritize functional correctness rather than exploring algorithm strategy's efficiency. 

Our study highlights several principles that should be considered when designing benchmarks for evaluating the performance of LLM-generated code. First, input scaling and diversity should be explicitly designed to expose algorithmic differences. Tasks should include sufficiently large and varied inputs so that implementations with different time or space complexities exhibit measurable performance gaps. Second, evaluation methods should incorporate robust measurement practices, such as repeated executions, statistical significance testing, and controlled runtime environments, to reduce measurement noise. Third, benchmarks should consider multiple performance dimensions beyond runtime, including memory consumption, scalability across increasing input sizes, and the stability of performance across runs. Finally, task design itself may also emphasize performance-sensitive workloads, such as scenarios in which different algorithmic strategies lead to substantially different asymptotic or constant-factor behavior.

While our work later demonstrates that enhanced test suites can reveal relatively more hidden efficiency differences, designing benchmarks that systematically incorporate these principles remains an open challenge. Without such advances, current benchmarks risk continuing to misrepresent the true performance of LLM-generated code.

\phead{Strengthening Test Suites Beyond Correctness.} 
Our findings in RQ2 show that existing benchmark test suites are primarily designed to validate functional correctness, with input sizes that are often too small to reveal meaningful performance differences. As a result, many performance improvements remain hidden under the default tests. This highlights the need for test generation approaches that explicitly target performance-critical execution paths. Our investigation in RQ3 further supports this finding. By incorporating more powerful test suites that are generated to stress the program, we can reveal an average of 32.66\% more performance differences between canonical and performant implementations.

While our multi-agent approach improves test diversity, it is limited by using LLM-based agents. In some cases, the generated tests may fail to exercise worst-case execution behavior due to the non-deterministic nature of LLMs, and may also misidentify the main function in programs with nested functions, leading to correctness issues.

Future work may focus on guided input generation that targets specific complexity patterns, such as inputs that trigger worst-case behavior or workloads that scale progressively to reveal asymptotic differences ~\citep{lemieux2018perffuzz,petsios2017slowfuzz}. In addition, agents could incorporate profiling feedback to identify hot execution paths and generate inputs that repeatedly exercise those paths.~\citep{llm4effi} Another promising direction is to adaptively refine test inputs, where the agent iteratively increases input size or structural complexity based on observed runtime differences until statistically significant performance gaps emerge. Finally, integrating static analysis or complexity estimation could help agents identify code regions that are likely to produce performance divergence and generate targeted stress tests accordingly. These directions are essential for enabling test suites that can reliably expose meaningful performance differences and support trustworthy evaluation of code efficiency. 

\phead{Beyond Function-Level Python Benchmarks.} 
Our study focuses on Python-based, function-level benchmark tasks, which provide a controlled setting for comparing implementations but may not fully capture real-world performance scenarios. In practice, performance issues often arise from interactions across multiple components, such as object lifecycles and repeated method calls within larger systems. As a result, class-level or repository-level benchmarks can provide a more realistic setting for evaluating performance, since they involve richer program structures, larger codebases, and more complex execution contexts where performance differences accumulate over time~\citep{swebench2024, repocoder2023, gorilla2024, repobench2024}.

In addition, real-world performance is often influenced by factors that are rarely represented in function-level tasks. For example, many applications are I/O-bound, where performance depends on file operations, database queries, or network communication rather than pure algorithmic computation~\citep{zhao2022large, li2019detecting}. Similarly, concurrent or asynchronous programs introduce scheduling and synchronization overheads that can significantly affect performance behavior ~\citep{selakovic2016performance, zhang2019understanding}. Another important scenario involves long-running workloads, such as batch data processing or server-side services, where small per-call efficiency improvements may accumulate into substantial runtime differences ~\citep{sweperf2025, zhang2024trace, eismann2022case}. Future benchmarks should therefore incorporate these more realistic workloads and system-level interactions to better evaluate the effectiveness of optimization techniques and LLM-generated performance improvements in practical software systems.
\section{Related Work}\label{sec:related}

\phead{Benchmarks for Code Generation}
Code-generation benchmarks evaluate LLMs at different levels of program scope. At the function level, benchmarks such as HumanEval~\citep{humaneval}, MBPP~\citep{mbpp}, APPS~\citep{apps2021}, DS-1000~\citep{ds10002023}, CodeContests~\citep{li2022competition}, LiveCodeBench~\citep{jain2025livecodebench}, and BigCodeBench~\citep{zhuo2025bigcodebench} assess whether a model can synthesize a function or program that satisfies an input-output specification and passes executable tests. Class-level benchmarks, including ClassEval~\citep{du2023classeval}, JavaBench~\citep{cao2024javabench}, and subsequent class-generation benchmarks~\citep{rahman2025beyond,chen2026classeval}, extend evaluation to multiple methods, class state, and interactions among class members. Repository-level benchmarks further evaluate code generation and software-engineering capabilities in the context of project structure, dependencies, and existing source files~\citep{swebench2024,repobench2024,li2024deveval,le2025impacts,li2025fea}. Across these levels, evaluation primarily emphasizes functional correctness.

However, functional correctness does not establish whether generated code is computationally efficient. Performance-oriented benchmarks such as EffiBench~\citep{effibench}, Enamel~\citep{enamel}, EvalPerf~\citep{evalperf}, and Mercury~\citep{mercury} evaluate runtime efficiency for Python functions. EffiBench-X~\citep{qing2026effibench} extends EffiBench to five additional languages: C++, Java, JavaScript, Ruby, and Golang. These benchmarks provide a canonical solution and a performant implementation for each task. The performant implementation serves as the runtime reference for evaluating LLM-generated code using executable tests and runtime measurements. In this paper, our study examines the four Python function-level performance benchmarks: EffiBench, Enamel, EvalPerf, and Mercury.

\phead{Empirical Studies on Code Efficiency}
A number of studies have explored methodologies for evaluating code efficiency~\citep{stoico2025empirical, guimaraes2025analyzing}. These include executing code in local environments~\citep{nofuneval2024,cheng2023revisiting}, using containerized execution for reproducibility~\citep{khan2024xcodeeval}, and designing performance-oriented test cases to stress input sizes and expose efficiency differences~\citep{enamel,evalperf,mercury,coffee25}. 
Other studies directly evaluate the efficiency of LLM-generated code. Runtime-based evaluations on HumanEval~\citep{humaneval}, MBPP~\citep{mbpp}, and LeetCodeEval show that functional correctness does not necessarily correlate with execution efficiency~\citep{niu2024evaluating}. Further empirical studies compare LLM-generated solutions with human-written implementations across runtime performance, memory usage, and energy consumption~\citep{coignion2024performance,islam2025evaluating, solovyeva2025ai}. 
These studies show growing interest in evaluating the computational efficiency of generated code. However, they mainly focus on assessing LLM-generated implementations, whereas our work re-examines the benchmarks themselves by testing whether benchmark-provided performant reference implementations are statistically distinguishable from their corresponding canonical solutions.

\phead{Automated Test Generation with LLMs}
Beyond code generation, recent work has investigated the use of large language models (LLMs) for automated test generation. A major line of work focuses on functional unit test generation, where LLMs are used to produce executable tests for a given method, function, class, or API under test~\citep{schafer2023empirical,yuan2024evaluating, chen2024chatunitest}. These studies primarily aim to improve functional correctness, compilability, and code coverage. Subsequent work further improves LLM-generated tests by incorporating execution feedback, coverage feedback, search-based exploration, or mutation testing signals~\citep{lemieux2023codamosa,alagarsamy2024a3test, altmayer2025coverup, pan2025aster}. In parallel, other studies examine supporting tasks for test generation, including assertion and oracle generation, test repair, test-suite augmentation, and model adaptation for unit testing~\citep{rao2023cat,shin2024domain,lukasczyk2022pynguin,wang2406testeval, hossain2025togll}.
Despite these advances, most LLM-based test generation techniques focus on correctness-oriented testing rather than performance-oriented testing. Recent work on performance-oriented test generation, such as WEDGE~\citep{wedge2025} and COFFE~\citep{coffee25}, has begun to address this gap by generating inputs that expose inefficient program behavior. Among them, COFFE is most closely aligned with our setting, as it studies stressful test generation for evaluating the efficiency of LLM-generated code at the Python function level. Our work complements this direction by examining how test-suite strength affects the reliability of performance evaluation.

\phead{LLM-based Code Generation}
Recent advances in large language models (LLMs), such as GPT~\citep{gpt4}, LLaMA~\citep{llama}, Gemini~\citep{gemini15}, Claude~\citep{anthropic}, DeepSeek~\citep{deepseek_chat_2025}, and Mixtral~\citep{mixtral}, have substantially improved automated code generation. Most existing work only focus on generating functionally correct code, where models are assessed by whether their generated programs pass the provided test cases or benchmark oracles~\citep{fenglinsoen2025, selfcodealign, codet5plus, alphacode}. Beyond correctness, recent studies have begun to investigate whether LLMs can generate computationally efficient code, focusing on runtime-aware prompting, optimization-oriented fine-tuning, self-improvement, and efficiency-guided code generation~\citep{effilearner,llm4effi, feng2024,peng2025perfcodegen, waghjale2024ecco}.
\section{Threats to Validity}
\label{sec:threats}

\phead{Internal Validity.}
One potential threat arises from the variability of runtime measurements in RQ1, RQ3 and RQ4, as execution time can be affected by system noise (e.g., OS scheduling and caching). To mitigate this, we execute each implementation 30 times and apply the Mann–Whitney U test to ensure that observed differences are statistically significant.
Another potential internal threat is the subjectivity in the manual analysis in RQ2. Since open coding relies on human judgment, bias may affect how code changes are interpreted and categorized. We mitigate this by having two authors independently label each task, resolving disagreements through discussion, and achieving a high inter-rater agreement (Cohen's Kappa: 0.93).

\phead{External Validity.}
Our study focuses on four performance benchmarks: EffiBench~\citep{effibench}, Enamel~\citep{enamel}, EvalPerf~\citep{evalperf}, and Mercury~\citep{mercury} which primarily consist of Python based programming tasks at the function level. As a result, our findings may not directly generalize to other programming languages, large scale software systems, or performance scenarios involving input-output operations, concurrency, or system level interactions. However the qualitative insights from running code multiple times with statistical testing and performance test cases are expected to generalize to similar benchmark studies. Researchers should thus consider the potential biases highlighted by our study when exploring similar datasets.

\phead{Construct Validity.} 
A construct threat concerns how we operationalize ``performance'' in Python function-level tasks. Our measurements focus on execution time, which fits the benchmark setting studied in this paper. However, in larger software systems, performance may also involve memory consumption, energy usage, throughput, latency, and interactions among components. Thus, our findings should be interpreted as evidence about function-level execution-time efficiency rather than a complete assessment of all performance dimensions. Future studies that move beyond function-level benchmarks should consider these additional performance dimensions when evaluating LLM-generated code.

\section{Conclusion}\label{sec:conclusion}

Existing Python function-level performance benchmarks are widely used to evaluate whether LLMs can generate efficient code. However, when these benchmarks report little or no execution-time difference between generated implementations and canonical solutions, it is unclear whether the result reflects a limitation of the LLMs or a limitation of the benchmark setting. In this paper, we revisit four benchmarks, EffiBench, Enamel, EvalPerf, and Mercury, using repeated execution and statistical testing. Across 1,538 tasks, only 6.11\% of benchmark-provided performant implementations are significantly faster than their corresponding canonical solutions under the original test suites. This result shows that the original benchmark settings often do not provide enough evidence to support reliable performance comparisons.

Our analysis further shows that non-significant results arise from two main causes. Some benchmark-provided performant implementations contain no meaningful performance change, such as refactoring or no code changes. Others contain algorithmic or data-structure changes that may improve runtime, but the original tests do not sufficiently stress the performance-critical paths where those changes matter. To address this limitation, we develop a multi-agent framework for generating deterministic performance-oriented tests. With stronger tests, more benchmark-provided performant implementations become statistically distinguishable from their canonical solutions, and LLM-generated implementations show statistically significant performance improvements in 22.19\% of the evaluated cases.

Overall, our findings show that function-level Python performance benchmarks need to be evaluated carefully before they are used to judge whether performant code is truly more efficient than a canonical solution. The evaluation should rely on multiple executions and statistical testing, rather than single-run or small-sample runtime comparisons. This direction is important for future benchmark design: researchers who build performance benchmarks should explicitly account for runtime noise and include statistical testing in the evaluation protocol.
\section{Declarations}

\section*{Funding}
This research received no external funding.

\section*{Ethical Approval}
Not applicable.

\section*{Informed Consent}
Not applicable.

\section*{Author Contributions}
The contributions of the authors are as follows.

Nhat Minh Le: Conceptualization, Methodology, Analysis, Investigation, Experiments, Writing - original draft, Writing - review and editing.

Yisen Xu: Analysis, Validation, Investigation, Writing - review and editing.

Zhijie Wang: Analysis, Validation, Investigation, Writing - review and editing.

Tse-Hsun (Peter) Chen: Conceptualization, Supervision, Funding acquisition, Validation, Writing - review and editing.

\section*{Conflict of Interest}
The authors declare that they have no conflict of interest.

\section*{Data and Code Availability Statement}
Our artifacts, including the replication package, all generated code, and measured code performance data, are publicly available online~\citep{artifact}.

\bibliographystyle{spbasic}
\bibliography{references}

\end{document}